\documentclass[acmsmall]{acmart}

\usepackage{soul}
\usepackage{color, multirow}
\usepackage{graphicx, subfigure}
\usepackage[utf8]{inputenc}

\setcopyright{acmcopyright}
\acmJournal{JETC}
\acmYear{2019} 
\acmVolume{1}
\acmNumber{1}
\acmArticle{1}
\acmMonth{1}
\acmPrice{15.00}

\setcopyright{acmcopyright}

\acmDOI{000000x.000000x}

\begin{document}


\title{Thread Batching for High-performance Energy-efficient GPU Memory Design}

	\author{Bing Li}
	\orcid{0000-0003-0732-2267}
	\affiliation{%
		\institution{Duke University}
		\department{Department of Electrical and Computer Engineering}
		\city{Durham}
		\state{NC}
		\postcode{27701}
		\country{USA}
	}
	\affiliation{%
		\institution{Army Research Office, Research Triangle Park}
		\country{USA}}
	\email{bing.li.ece@duke.edu}
	\author{Mengjie Mao}
	\affiliation{%
			\institution{MathWorks Inc.} 
	\country{USA}}
\email{meng.j.mao@gmail.com}

\author{Xiaoxiao Liu}
\affiliation{%
	\institution{AMD}
	\country{USA}
}
\email{xil116@pitt.edu}

\author{Tao Liu}
\affiliation{%
	\institution{Florida International University} 
	\department{Department of Electrical and Computer Engineering}
	\city{Miami}
	\state{FL}
	\postcode{33174}
	\country{USA}}
	\email{tliu023@fiu.edu}
\author{Zihao Liu}
\affiliation{%
	\institution{Florida International University} 
	\department{Department of Electrical and Computer Engineering}
	\city{Miami}
	\state{FL}
	\postcode{33174}
	\country{USA}}
	\email{zliu021@fiu.edu}
	
\author{Wujie Wen}
\affiliation{%
	\institution{Florida International University} 
	\department{Department of Electrical and Computer Engineering}
	\city{Miami}
	\state{FL}
	\postcode{33174}
	\country{USA}}
\email{wwen@fiu.edu}

\author{Yiran Chen}
	\affiliation{%
	\institution{Duke University}
	\department{Department of Electrical and Computer Engineering}
	\city{Durham}
	\state{NC}
	\postcode{27701}
	\country{USA}
}
\email{yiran.chen@duke.edu}
\author{Hai (Helen) Li}
	\affiliation{%
	\institution{Duke University}
	\department{Department of Electrical and Computer Engineering}
	\city{Durham}
	\state{NC}
	\postcode{27701}
	\country{USA}
}
\email{hai.li@duke.edu}

\renewcommand\shortauthors{Li,B. et al}

\thanks{
	This work was supported by US Department of Energy (DOE) under Grant SC0017030. Bing Li is supported by the NRC Associate Fellowship Award. This work is an extended version of a paper that was published at the Proceedings of Design Automation Conference (DAC) 2016, entitled "TEMP: Thread Batch Enabled Memory Partitioning for GPU". 
}
\authorsaddresses{
	Author's addresses: B. Li, Y. Chen and H. Li are with the Department of Electrical and Computer Engineering, Duke University, Durham, NC 27708 USA (e-mail: bing.li.ece@duke.edu, yiran.chen@duke.edu, hai.li@duke.edu).
	M. Mao received Ph.D. from University of Pittsburgh in 2016. He is currently a senior software engineer in the MathWorks Inc. (e-mail: meng.j.mao@gmail.com).
	X. Liu received his Ph.D. from University of Pittsburgh in 2016. She is currently a member of technical staff at AMD (e-mail: xil116@pitt.edu).
	T. Liu, Z. Liu and W. Wen are with the Department of Electrical and Computer Engineering, Florida International University, Miami, FL 33174 USA (e-mail: tliu023@fiu.edu, zliu021@fiu.edu, wwen@fiu.edu).
}
\thanks{}
\begin{abstract}
	Massive multi-threading in GPU imposes tremendous pressure on memory subsystems. 
	Due to rapid growth in thread-level parallelism of GPU and slowly improved peak memory bandwidth, memory becomes a bottleneck of GPU's performance and energy efficiency.
    In this work, we propose an integrated architectural scheme to optimize the memory accesses and therefore boost the performance and energy efficiency of GPU.
	Firstly, we propose a \emph{thread batch enabled memory partitioning (TEMP)} to improve GPU memory access parallelism. In particular, TEMP groups multiple thread blocks that share the same set of pages into a thread batch and applies a page coloring mechanism to bound each \textit{stream multiprocessor (SM)} to the dedicated memory banks. After that, TEMP dispatches the thread batch to an \textit{SM} to ensure high-parallel memory-access streaming from the different thread blocks.
	Secondly, a \emph{thread batch-aware scheduling (TBAS)} scheme is introduced to improve the GPU memory access locality and to reduce the contention on memory controllers and interconnection networks.
	Experimental results show that the integration of TEMP and TBAS can achieve up to 10.3\% performance improvement and 11.3\% DRAM energy reduction across diverse GPU applications.
We also evaluate the performance interference of the mixed CPU+GPU workloads when they are run on a heterogeneous system that employs our proposed schemes. Our results show that a simple solution can effectively ensure the efficient execution of both GPU and CPU applications.

\end{abstract}



\maketitle

\section{Introduction}
\label{sec:intro}

The use of Graphics Processing Units (GPUs) has been extended from fixed graphics acceleration to general purpose computing, including image processing, computer vision, machine learning, and scientific computing. 
GPU is widely employed in various platforms ranging from embedded systems to high-performance computing systems~\cite{shimpi2012inside}.

GPU heavily relies on massive threading to achieve high throughput. 
However, it commonly incurs intensive memory accesses, which may limit the performance and energy efficiency of GPU~\cite{jog2013orchestrated} as the result of the high overhead of device memory access\footnote{In this work, we use device memory and memory interchangeably.}.
Though large-capacity and low-overhead cache have been adopted by GPU to alleviate the impact of inefficient memory accesses~\cite{abdel2013warped,mao2014exploration}, the available cache per thread is far below the demand of most GPU applications~\cite{jia2012characterizing}.
The pressures on device memory, \textit{i.e.}, DRAMs, in GPU are still severe.

Memory scheduling is one of the primary architectural techniques to improve memory efficiency as it is able to optimize the memory access parallelism and locality in multi-core systems~\cite{mutlu2007stall, mutlu2008parallelism, kim2010thread, ebrahimi2011parallel, usui2016dash}.
However, the existing memory scheduling algorithms are usually associated with expensive implementation~\cite{liu2012software} and also insufficient to handle the intensive memory accesses in GPU~\cite{yuan2009complexity, ausavarungnirun2012staged}.

The \emph{memory partitioning (MP)} based on operating system (OS) memory management is another viable approach to improve memory efficiency and reduce inter-thread memory interference.
Memory partitioning generally divides memory resources and assigns them to threads, and every thread accesses its exclusive memory space~\cite{mi2010software,liu2012software,jeong2012balancing,xieimproving,suzuki2013coordinated}.
Memory partitioning is promising to improve memory efficiency in the GPU system because of the following reasons:
1)~The memory address space in the heterogeneous system is pageable. The memory pages can be allocated to the GPU threads by OS; and
2)~The threads in GPU are nearly homogeneous. When they are evenly dispatched to \textit{stream multiprocessors} (SMs), the fairness and parallelism of their access to memory can be guaranteed. This assertion, however, may be invalid in other multi-core systems due to the disparity of memory bandwidth required by their threads~\cite{xieimproving}.

Unfortunately, the existing memory partitioning mechanisms for multi-core systems cannot straightly applied to the GPU.
For instance, the \emph{memory bank partitioning (MBP)}~\cite{mi2010software,jeong2012balancing,liu2012software,xieimproving}, which allows each thread to access the exclusive memory bank.
MBP aims for the multi-program systems which have few parallel threads.
Differently, GPU always runs massive threads, of which the number is orders of magnitude larger than the available banks.
It is impossible to allow every thread has the exclusive memory bank.
Moreover, all threads in a GPU application share an unified address space~\cite{fermi}.
Their memory accesses interweave together and is difficult to be separated by using memory partitioning technique.

To address the above problems, we propose an integrated solution to improve the performance and energy efficiency of GPU applications.
The integrated solution is composed of the \emph{thread batch enabled memory partitioning} (TEMP) to enhancing the memory access parallelism and the \emph{thread batch-aware scheduling} (TBAS) to improve the memory access locality.
Specifically, TEMP assigns the majority of memory requests from the same SM to the dedicated banks to ensure the parallelism of memory accesses of threads.
The thread blocks that share the same set of pages are grouped into a thread batch and then are dispatched to an SM as a whole.
Meanwhile, by applying the page coloring mechanism, the accessed pages are mapped to the dedicated banks which are associated with the same SM~\cite{lin2008gaining}. 
In this way, TEMP minimizes the interference of memory accesses from different SMs and improves the parallelism of memory accesses. 
Moreover, TBAS prioritizes the execution of thread batches to preserve the locality of memory accesses.
The thread batches that access the same row in one bank are clustered and scheduled together.  
Accordingly, TBAS effectively alleviates the contention on the memory controllers and the congestion on the reply network connecting the memory partitions to SMs.
%

We compare TEMP and TBAS with some representative thread scheduling techniques, including the cache-conscious wavefront scheduler (CCWS)~\cite{rogers2012cache}, OWL~\cite{jog2013owl} and the bandwidth-aware policy (BW-AWARE)~\cite{agarwal2015page}. We set CCWS as our baseline and integrate OWL, BW-AWARE and our techniques on top of CCWS.
The benchmarks consist of not only GPU applications but also the combined CPU-GPU applications.
Experimental results show that after applying TEMP and TBAS, the GPU system achieves 10.3\% performance improvement and 11.3\% reduction of the DRAM energy consumption for the evaluated GPU applications
compared to the baseline. 
The results of the combined CPU-GPU workloads demonstrate that a simple yet effective solution is capable of addressing the interference incurred by the CPU executions and ensuring high execution efficiency in GPU applications using TEMP and TBAS with negligible performance degradation on the CPU side.

The rest of this paper is organized as follows:
Section~\ref{sec:bg} introduces the background of the heterogeneous CPU-GPU system and memory system;
Section~\ref{sec:tbmp} and Section~\ref{sec:tbas} describe the details of TEMP and TBAS, respectively;
Section~\ref{sec:setup} summarizes our experimental setup;
Section~\ref{sec:re} presents the experimental results and related analyses;
Section~\ref{sec:rw} discusses the related works;
and Section~\ref{sec:cc} concludes our work.
\section{Background}
\label{sec:bg}

\begin{figure}[t]
	\center
	\includegraphics[width=.8\columnwidth]{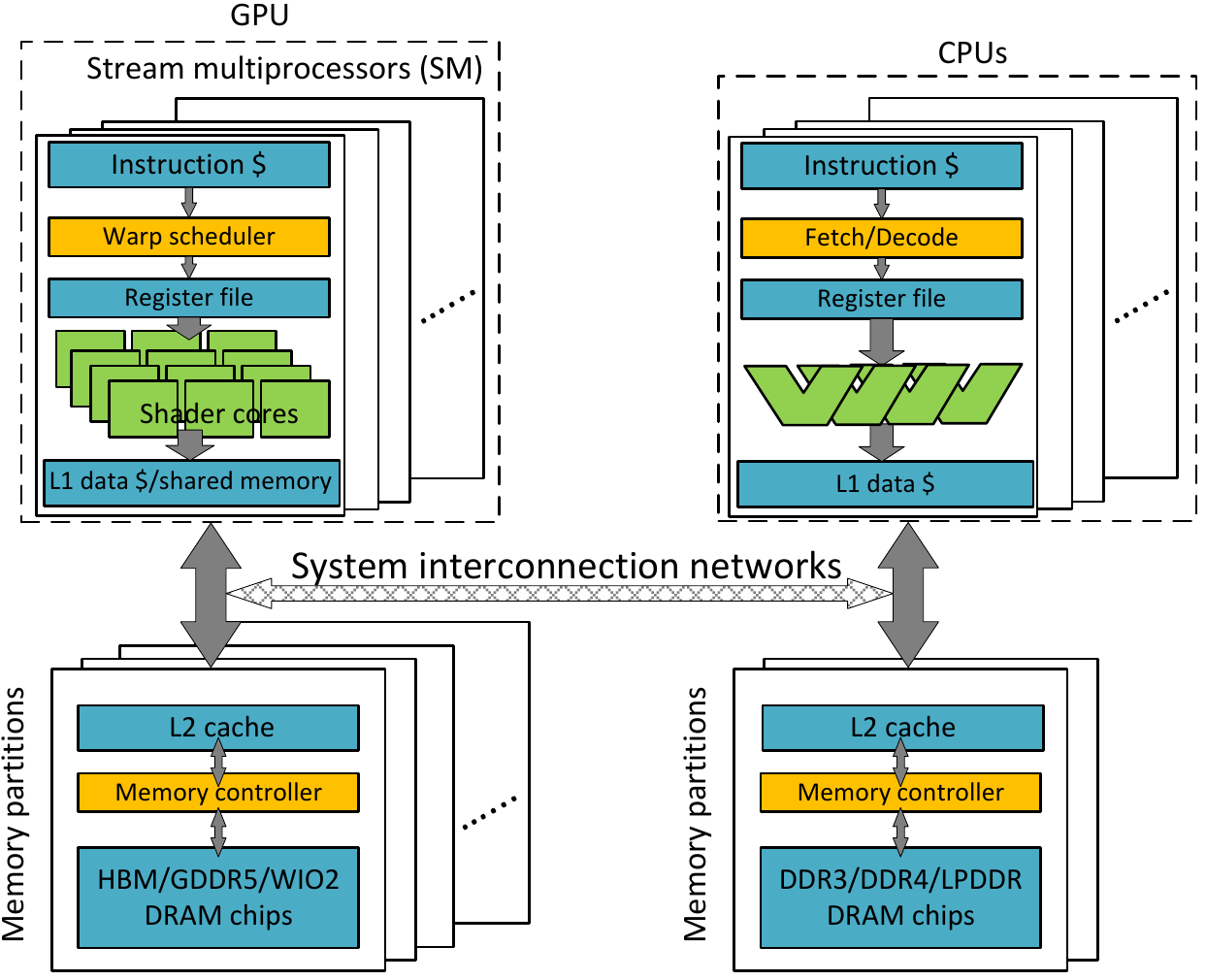}
	\caption{Organization of a heterogeneous CC-NUMA system.}
	\Description[Fig. 1]{Organization of a heterogeneous CC-NUMA system.}
	\label{fig:heter_arch}
\end{figure}

The heterogeneous CPU-GPU integrated systems are evolving towards unified memory address space~\cite{chu2013amd}. 
Because of discrepant bandwidth requirements, it is anticipated that GPU will be still physically attached with bandwidth-optimized DRAM, while CPU is attached with capacity- and cost-optimized DRAM.
DRAMs of GPU and CPU share a unified memory address space~\cite{agarwal2015page}. 
In such heterogeneous \emph{cache coherent non-uniform memory access} (CC-NUMA) system, a computing unit has different access delays to local and remote memories even it sees a unified address space. 
Fig.~\ref{fig:heter_arch} shows a heterogeneous CC-NUMA system including several CPUs and a GPU. 
The system interconnection networks bridge two memories and maintain the coherence between caches of the CPUs and the GPU.

Heterogeneous CC-NUMA allows better programmability and finer-grained memory management of the GPU. OS can allocate the GPU pages in all memories. 
In this work, we use the default NUMA page placement policy in Linux, \textit{i.e.}, \emph{local}, which places as many pages as possible in the local memory. By using \texttt{local} policy we can avoid most bandwidth contentions between the CPUs and the GPU in heterogeneous CC-NUMA.

GPU programming models such as CUDA~\cite{cuda} and OpenCL~\cite{opencl} define the workload offloaded to a GPU as a \emph{kernel}.
A kernel is highly multi-threaded where all the threads are encapsulated in a \emph{grid}.
Within a grid, the threads are partitioned into three-dimensional thread blocks, each of which contains up to thousands of threads.
During executions, each thread block is dispatched as a whole to a SM. Every SM holds a complete single instruction multiple data (SIMD) pipeline. 
Each thread block in the SM is further partitioned into many fixed-size \emph{warps} that are atomically scheduled by a \emph{warp scheduler} and executed in the SIMD fashion.
The L2 caches of CPUs and of GPU are separated and placed in different memory partitions, each of which has its own memory channel.
The on-chip caches, including L1 data and instruction caches in the CPUs and the GPU, are connected to the L2 caches via a mesh network.
In such design, GPU can use page-fault memory rather than be restricted to page-locked memory~\cite{branover2012amd} and non-pageable memory~\cite{fermi}. 

\subsection{Heterogeneous CC-NUMA}

\subsection{DRAM Basics}

\begin{figure}[t]
\center
\includegraphics[width=.8\columnwidth]{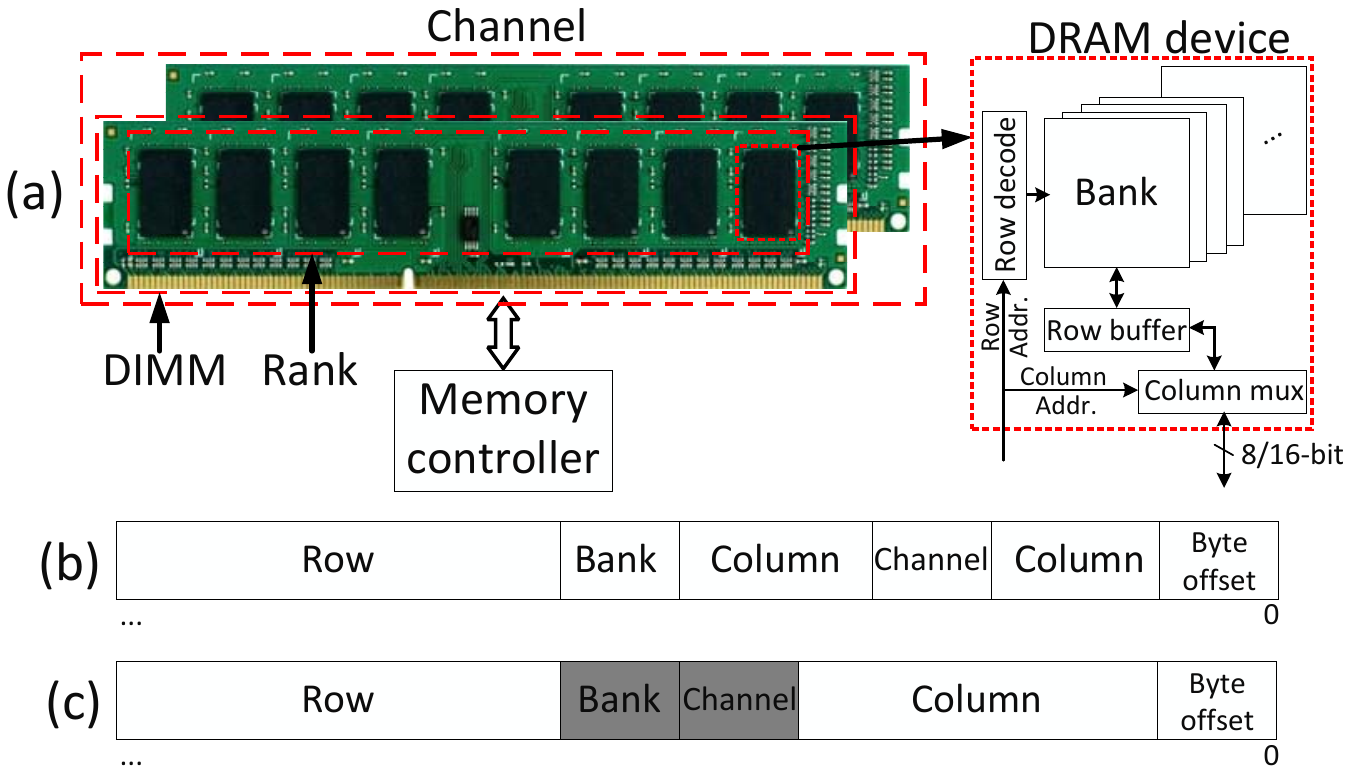}
\caption{(a) The organization of a DRAM channel; (b) The default DRAM address mapping used in this work; (c) The modified DRAM address mapping used for page coloring. The shadow segment is the color bits.}
\Description[Fig.2]{(a) The organization of a DRAM channel; (b) The default DRAM address mapping used in this work; (c) The modified DRAM address mapping used for page coloring. The shadow segment is the color bits.}
\label{fig:DRAM}
\end{figure}

A modern JEDEC compliant DDRx DRAM system consists of one or more channels, each of which has its own data buses, command buses, and address transferring.
Fig.~\ref{fig:DRAM}(a) depicts the basic organization of a DRAM channel, which also has a \emph{memory controller} (MC) to control the operations on the channel.
A channel may include multiple DIMMs.
Within each DIMM, there are several ranks, each of which consists of multiple DRAM devices.
In DDR3, a DRAM device contains eight banks.
The data of each bank are always pre-loaded to its private row buffer before being accessed.

DRAM address mapping complies with the DRAM organization.
The address mapping scheme in Fig.~\ref{fig:DRAM}(b)~\cite{bakhoda2009analyzing} is the baseline DRAM address mapping we used in our heterogeneous architecture.
The address mapping scheme in Fig.~\ref{fig:DRAM}(c) is used for the page coloring mechanism in our work.
If the number of page offset bits is not greater than the sum of the column and byte offset bits, by using page coloring, a GPU page can be mapped to arbitrary channel, rank, bank or row in a bank.

The memory usage efficiency is mainly determined by \emph{bank-level parallelism} (BLP)~\cite{mutlu2008parallelism} and row locality measured by \emph{row buffer hit rate} (RBHR). 
All the banks in a DRAM can be accessed concurrently as each bank has its own address decoder and sensing logic.
However, only one bank can put/receive the data on/from the shared bus at a time.
All memory requests (reads and writes) need to go through the row buffer.
Memory access latency and energy can be reduced when the access hits on the row buffer as no row activation is needed.
In multi-core systems, a variety of memory schedulers~\cite{mutlu2007stall,mutlu2008parallelism} have been proposed to improve the BLP and row locality as well as maximize the access fairness.
However, these designs are generally insufficient to handle the massive parallel memory requests of GPU~\cite{yuan2009complexity}.
In this work, we propose TEMP and TBAS to improve the DRAM efficiency in GPU by minimizing the inter-SM interference of memory accesses, which is the root reason of low BLP and low row locality of DRAM accesses~\cite{jeong2012balancing}.

\section{Thread Batch Enabled Memory Partitioning (TEMP)}
\label{sec:tbmp}

A na\"ive GPU memory partitioning may bind each SM to one or more banks. All the pages accessed by a thread block can be placed to the banks bound to the SM where the thread block is executed. Ideally, if there is no shared page among different thread blocks, the banks can be exclusively accessed by the associated SM. 
Unfortunately, page sharing between thread blocks commonly exists in GPU kernels. 
The simple page placement mentioned above is unable to separate the memory access streams raised from different SMs. 
To address the issue, we propose TEMP which \emph{identifies and forms} the thread blocks sharing pages (Section~\ref{subsec:tbf}) and \emph{dispatches} them to the same SM (Section~\ref{subsec:tbd}) so as to minimize the inter-SM interference of memory accesses. 
The group of these thread blocks sharing pages is noted as a \emph{thread batch}. The rest of this section will detail the design and implementation of TEMP.

\subsection{Thread Batch Formation}
\label{subsec:tbf}

\begin{figure}[t]
\center
\includegraphics[width=.8\columnwidth]{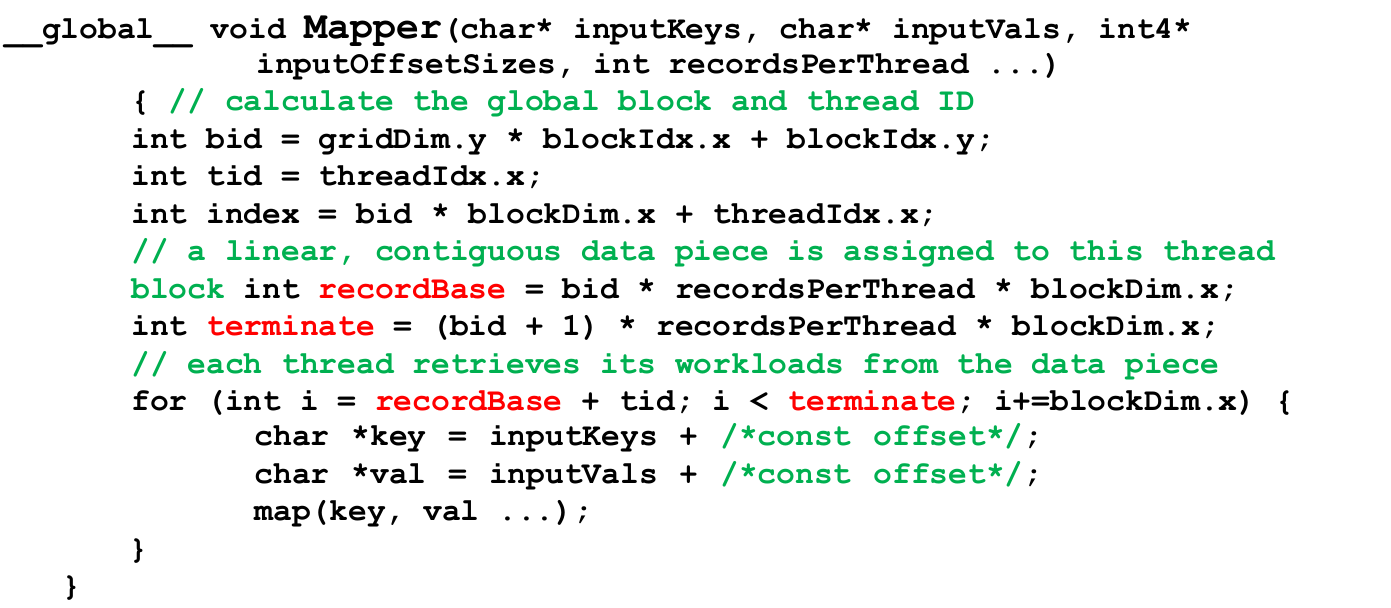}
\caption{Annotated code snippet of \texttt{Mapper} kernel in Mars library.}
\Description[Fig. 3.]{Annotated code snippet of \texttt{Mapper} kernel in Mars library.}
\label{fig:data_layout_code}
\end{figure}

By profiling the prevalent GPU benchmark suites, we observed there was two major types of thread-data mappings with some page sharing patterns in thread blocks
\footnote{In this work, we only consider the kernels constructed with 1D and 2D thread block/grid, because none of the profiled benchmarks employs 3D thread block/grid (see Table~\ref{tbl:app}).}.
The first type of thread-data mappings is: the data accessed by each thread block is clustered over a sequential address space. Fig.~\ref{fig:data_layout_code} shows the skeleton of the \texttt{Mapper} kernel in MapReduce engine of Mars~\cite{he2008mars}. This kernel employs fixed 1D thread blocks and scatters them to 1D or 2D grid. Generally, consecutive thread blocks sequentially access the 1D vector \texttt{inputKeys}, and each thread block accesses a linear address space ranging from \texttt{recordBase} to \texttt{terminate} within \texttt{inputKeys}.

\begin{figure}[t] 
	\center
	\includegraphics[width=.7\columnwidth]{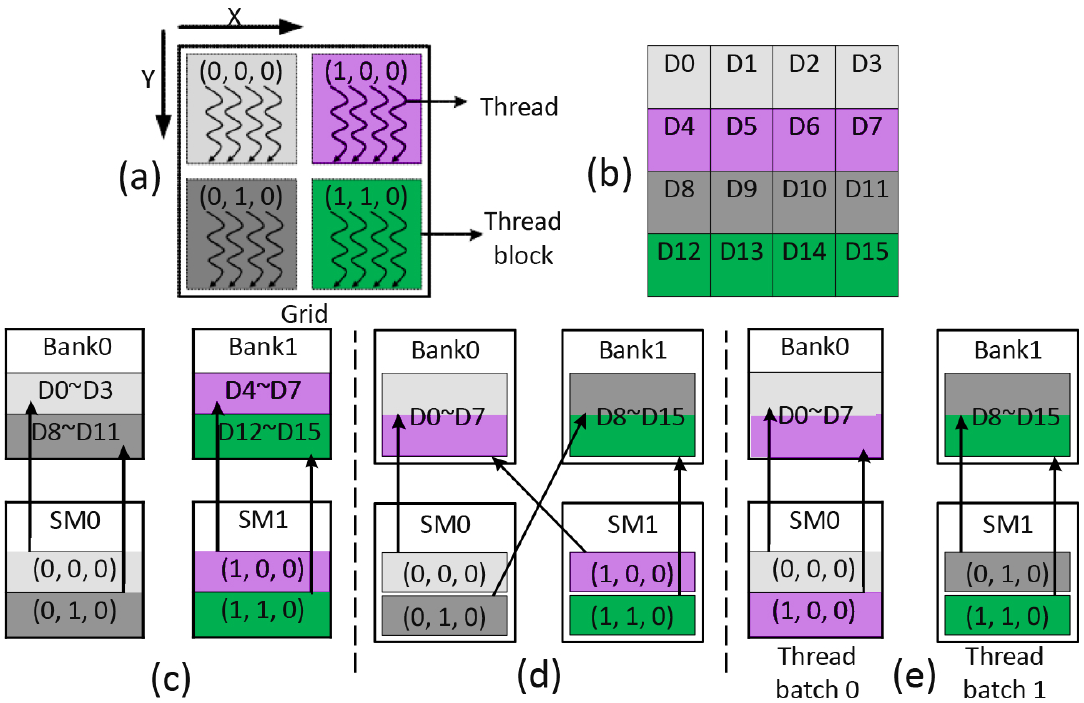}
	\caption{(a) A 2D grid -- 1D thread blocks; (b) The accessed data matrix; The memory access footprint when the page size is  (c) a matrix row, (d) two matrix rows without or (e) with thread batching.}
	\Description[Fig. 4.]{(a) A 2D grid -- 1D thread blocks; (b) The accessed data matrix; The memory access footprint when the page size is  (c) a matrix row, (d) two matrix rows without or (e) with thread batching.}
	\label{fig:data_layout}
\end{figure}
Fig.~\ref{fig:data_layout} simplifies and visualizes the first type of thread-data mapping. In this example we assume the grid of the kernel contains four thread blocks, each of which consists of four threads. 
The 1D thread blocks are arranged in a 2D grid. Their accessed data matrix is 
shown in Fig.~\ref{fig:data_layout}(b).
In this example, the first row of the data matrix is accessed by thread block (0,0,0), the second row is accessed by thread block (1,0,0), and so on. 
If the row address of the matrix aligns to a page, the SM-level page coloring can perfectly place the pages accessed by a SM to the bounded banks, as depicted in Fig.~\ref{fig:data_layout}(c). Here a page is equal to a matrix row.
However, if a page is composed of multiple matrix rows, say, two matrix rows, conventional thread block dispatching which interleaves thread blocks across SMs will generate interweaved memory accesses,
as shown in Fig.~\ref{fig:data_layout}(d).
In order to address the situation, we can pack those thread blocks accessing the same set of pages into a \emph{thread batch} and then dispatch the thread batch as a whole to a SM.
For the example shown in Fig.~\ref{fig:data_layout}(d), the 4 thread blocks can be grouped into 2 thread batches, each of which goes to a SM.
The memory accesses to banks 0 and 1 are successfully separated, as illustrated in Fig.~\ref{fig:data_layout}(e).

\begin{figure}[b]
\center
\includegraphics[width=.8\columnwidth]{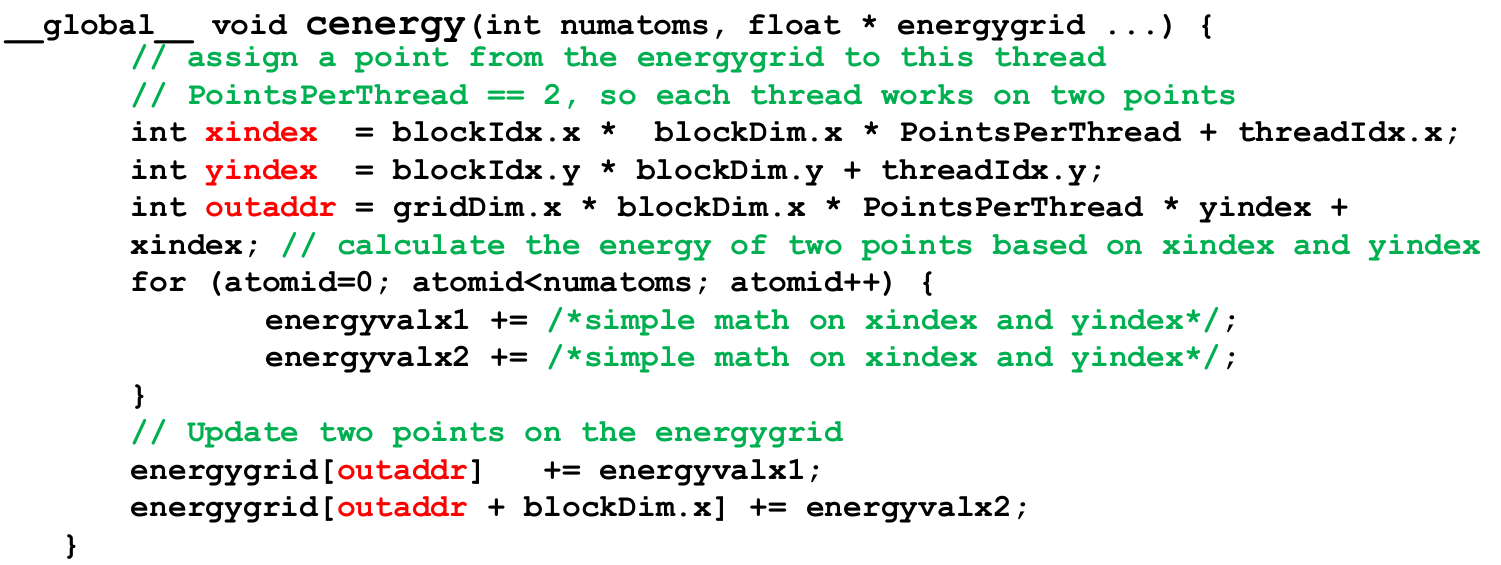}
\caption{Annotated code snippet of \texttt{cenergy} kernel in \texttt{CUTCP}.}
\Description[Fig. 5.]{Annotated code snippet of \texttt{cenergy} kernel in \texttt{CUTCP}.}
\label{fig:data_layout_2_code}
\end{figure}

\begin{figure}[b]
\center
\includegraphics[width=.7\columnwidth]{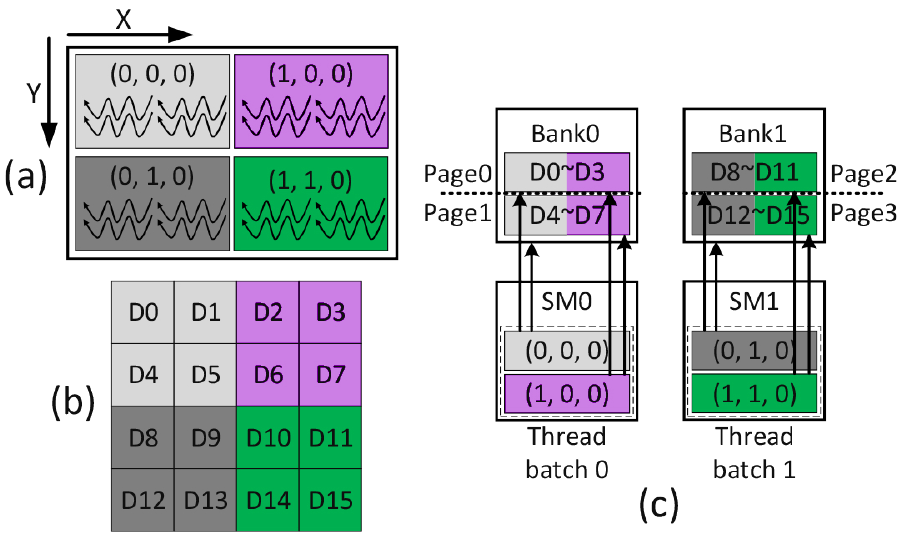}
\caption{(a) A 2D grid -- 2D thread blocks; (b) The accessed data matrix; (c) The memory access footprint with thread batching.}
\Description[Fig. 6.]{(a) A 2D grid -- 2D thread blocks; (b) The accessed data matrix; (c) The memory access footprint with thread batching.}
\label{fig:data_layout_2}
\end{figure}

The second type of thread-data mappings is that the data accessed by consecutive thread blocks are interleaved over a linear address space. 
Fig.~\ref{fig:data_layout_2_code} shows the code snippet of the \texttt{cenergy} kernel in the \texttt{CUTCP} benchmark~\cite{stratton2012parboil}. \texttt{CUTCP} computes the coulombic potential at a molecular grid \texttt{energygrid}. 
A point in \texttt{energygrid} is indexed by \texttt{xindex} and \texttt{yindex} generated from a thread's indexes. 
All threads form a 2D grid which is further tiled with 2D thread blocks.
Fig.~\ref{fig:data_layout_2} demonstrates 
a simplified thread-data mapping in this 2D grid.
The thread organization and accessed data matrix can be found in Fig.~\ref{fig:data_layout_2}(a) and (b), respectively. Here, we again assume one grid has four thread blocks, and each thread block has four threads.
In this example, every thread block has two active dimensions ($x$-axis and $y$-axis).
Each matrix row is accessed by two thread blocks while each thread block accesses two rows. 
In such a situation, the consecutive thread blocks likely access the same set of pages.
Similarly, we can pack those thread blocks sharing the same set of pages into one thread batch. 
Fig.~\ref{fig:data_layout_2}(c) gives a thread batching example where every matrix row in Fig.~\ref{fig:data_layout_2}(b) exactly forms one page. 
Thread blocks (0,0,0) and (1,0,0) share pages 0 and 1, while thread blocks (0,1,0) and (1,1,0) share pages 2 and 3.
Consequently, we can group thread blocks (0,0,0) and (1,0,0) into thread batch 0 and thread blocks (0,1,0) and (1,1,0) into thread batch 1.
By allocating pages 0 \& 1 into bank 0 and pages 2 \& 3 into bank 1, the memory accesses from SM 0 to bank 0 and from SM 1 to bank 1 are separated.

Those two major thread-data mapping scenarios indicate consecutive thread blocks may share pages. 
Accordingly, we introduce the \emph{thread block stride} to indicate the number of the consecutive thread blocks that belong to the same thread batch.
In the examples in Fig.~\ref{fig:data_layout}(c) and~\ref{fig:data_layout_2}(c), the thread block stride is 1 and 2, respectively.

To find the thread block stride of a GPU kernel, we profile a kernel given a page size at the compile time when the programmer determines the thread hierarchy and how the threads access the data matrices. 

\begin{figure}[b]
	\center
	\includegraphics[width=.7\columnwidth]{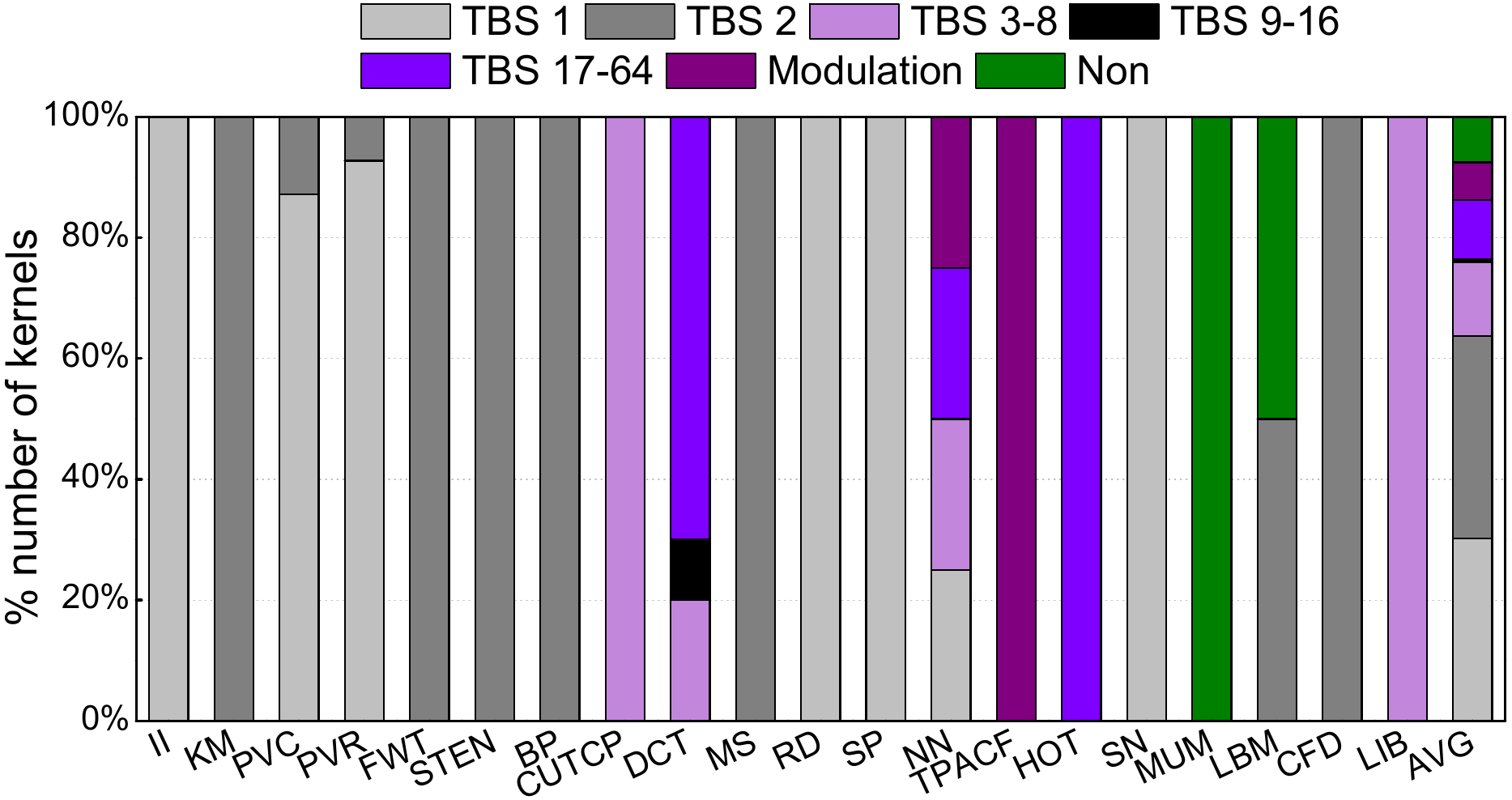}
	\caption{The distribution of thread block stride.}
	\Description[Fig. 7.]{The distribution of thread block stride.}
	\label{fig:kernel_dist}
\end{figure}

At the profiling stage, the start addresses of data matrices are set to zero. During dynamic memory allocation, the start memory address of a data matrix align to the beginning of the pages to guarantee the thread block stride to be found in the compile time. Fig.~\ref{fig:kernel_dist} shows the optimal thread block stride of some GPU applications. Optimal thread block stride denotes the thread block stride suppressing the most cross-batch page sharing.
Here, the page size is 4KB supported by most of the computer systems.
89\% of kernels achieve the minimum inter-thread batch page sharing through a batch formation with a fixed thread block stride.
There are also 6\% of kernels where the batch formation can be realized using modulation. 
Some kernels in \texttt{MUM} and \texttt{LBM} cannot be fitted with a formula for the batch formation.

The static compile-time profiling is sub-optimal since it cannot proactively remove the cross-batch page sharing. For example, the last thread block in a thread batch may share a page with the first thread block in its following thread batch, if those thread batches are formed with a fixed thread block stride.
In the next section we introduce a simple dynamic hardware approach which can support thread batching better relative to the static profiling.

\begin{figure}[t]
	\center
	\includegraphics[width=.7\columnwidth]{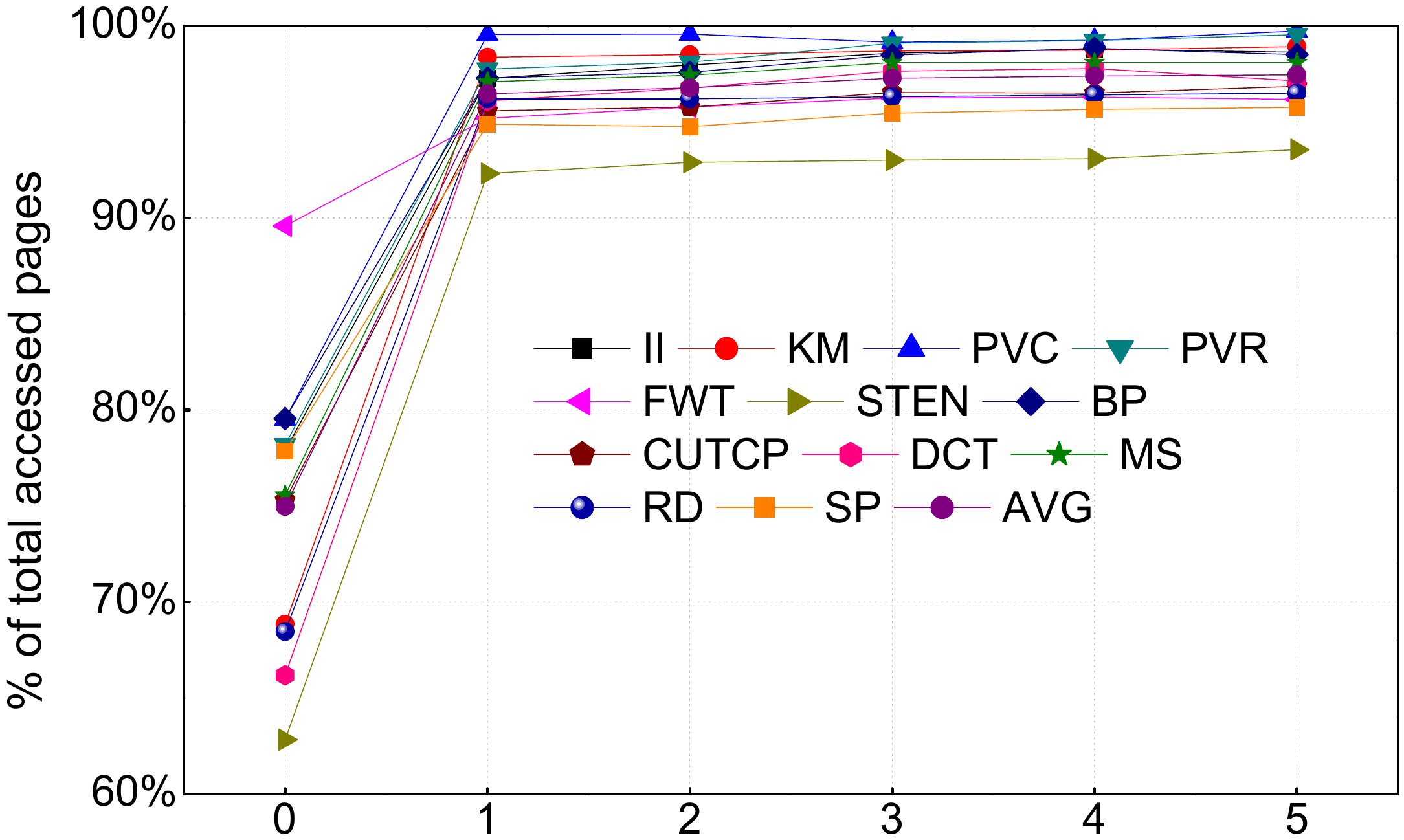}
	\caption{The accumulated percentage of cross-batch sharing between thread batches.}
	\Description[Fig. 8.]{The accumulated percentage of cross-batch sharing between thread batches.}
	\label{fig:shared_pages}
\end{figure}

We further analyze some GPU applications which form thread batches with the fixed thread block stride. The accumulated percentage of the pages shared by different sizes of consecutive thread batches is shown in Fig.~\ref{fig:shared_pages}. Horizontal axis shows the maximal distance of the shared pages among the thread batches.
Among all the accessed pages, nearly 75\% on average is exclusively accessed by a single thread batch and 22\% is accessed by two consecutive thread batches.
These two cases dominate the page access patterns in the thread batches ($>97\%$). There are more than 2\% of pages globally shared among all the thread batches in a kernel, such as program text pages.

\subsection{Serial Thread Block Dispatching}
\label{subsec:tbd}

Given that the thread batching and the cross-batch page sharing dominate the GPU applications, we propose \emph{serial thread block dispatching}. The consecutive thread blocks, which are very likely enclosed by the consecutive thread batches, are emitted to a SM. As such most thread batches are formed implicitly by the serial thread block dispatching, and most cross-batch page sharing are constrained within a SM. Now the cross-batch page sharing only happens when some thread blocks of a thread batch are distributed to multiple SMs. This would happen in the first and the last thread batch in an SM.

Traditional interleaved thread block dispatching, \textit{e.g.}, \emph{GigaThread engine} in NVIDIA GPU~\cite{fermi}, generates and dispatches a new thread block to an SM once the SM has an idle slot. Typically, the dispatching unit only passes the id of the new thread block to the SM, and the SM will construct a whole thread block according to the received thread block id. The dispatching unit generates the thread block ids sequentially and the thread block ids are dispatched to SMs randomly. To implement the deterministic and serial thread block dispatching, we introduce a dispatch queue in each SM.
The content, \textit{i.e.}, the thread block ids, in the dispatch queue are inserted before launching a kernel. Each SM receives similar amount of thread block ids in consideration of workload balance, which can be determined at the compile time.
During the kernel execution, the thread block ids are popped from the dispatch queue and emitted to the associated SM.

Compared to the traditional thread block dispatching, serial thread block dispatching avoids the stall of the launch of thread blocks. An SM can always pop a thread block id from its dispatch queue once it has an idle slot.
The implementation of the dispatch queue can be highly efficient since each SM only needs two extra registers to record the head and the tail of thread block ids.
The head register increments by one once a new thread block id (the head register itself) is popped. 
The dispatching of the thread block ends when the head register meets the tail of the thread block id stored in the second register. 
Thus, the serial thread block dispatching incurs marginal run-time and hardware overheads.
\section{Thread Batch-aware Scheduling (TBAS)}
\label{sec:tbas}

\begin{figure}[t]
\centering
\includegraphics[width=1\textwidth]{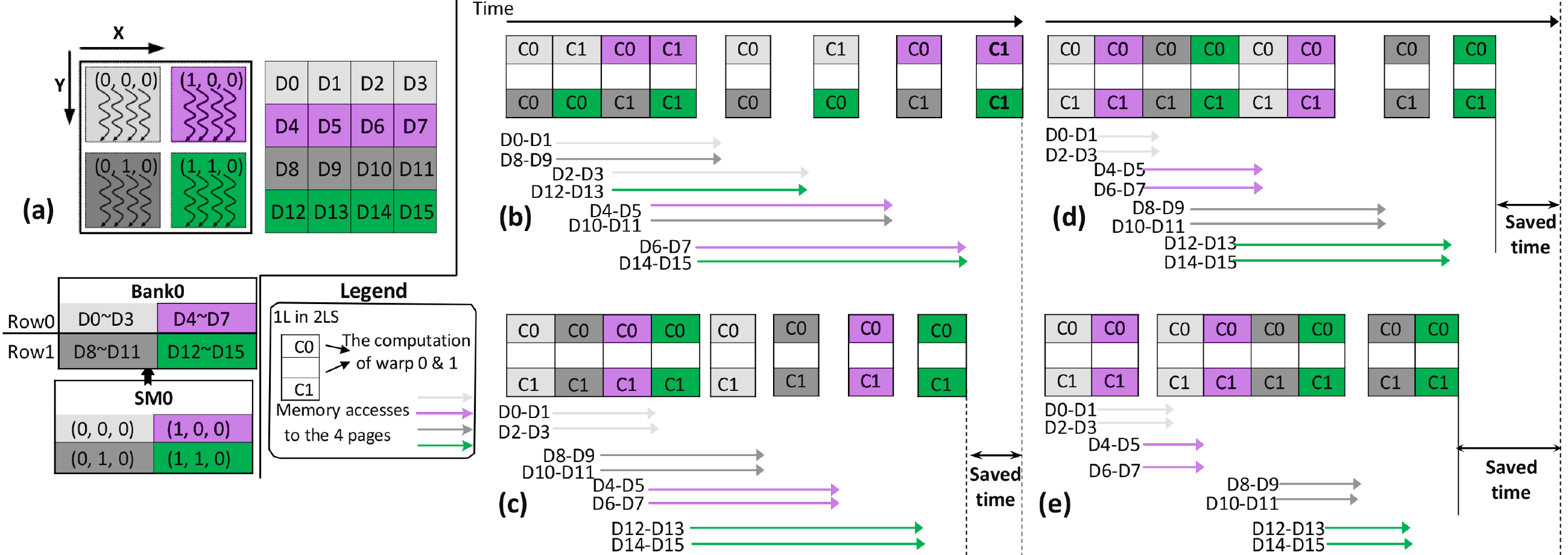}
\caption{(a) The thread organization, data matrix, thread batches, and memory access footprint of a kernel running on SM0; (b) The execution sequence of warps generated by CCWS; (c)--(e) The progressive improvement of TBAS. }
\Description[Fig. 9.]{(a) The thread organization, data matrix, thread batches, and memory access footprint of a kernel running on SM0; (b) The execution sequence of warps generated by CCWS; (c)--(e) The progressive improvement of TBAS.}
\label{fig:tbas}
\end{figure}

TEMP constrains the memory accesses from a SM within the associated memory banks, offering an opportunity to improve intra-bank/row locality by scheduling the execution of threads.
Accordingly, we propose TBAS that can be explained using the example in Fig.~\ref{fig:tbas}.

Fig.~\ref{fig:tbas}(a) presents the thread organization and the data matrix in the example.
In a GPU, there is only one SM (\textit{i.e.}, SM0) associated with its own DRAM bank.
Four thread batches, each of which consists of only one thread block, are formed and dispatched to SM0.
Every thread batch exclusively accesses its own page while the page layout of SM0's bank is also shown in Fig.~\ref{fig:tbas}(a).
We assume two pages are included in one row in the bank\footnote{Generally, the row size of a DRAM is multiple times greater than the smallest page size that the OS can support.}.
Every two threads in a thread block forms a warp.
Since there are four threads in one thread block, each thread block has two warps and total eight warps (or four thread blocks) are running on SM0.

Fig.~\ref{fig:tbas}(b) shows the execution of SM0 with a \emph{cache-conscious wavefront scheduler} (CCWS)~\cite{rogers2012cache}. CCWS was designed for improving the L1 cache locality in GPUs. It captures the intra-warp locality and decreases the L1 thrashing by limiting the number of active warps in a SM based on the L1 eviction information. 
Typically, CCWS only keeps a subset of warps running in SMs and throttles the rest of warps pending in the same SM if the cache thrashing is detected.
Once a warp in the running set encounters a stall, it will be demoted to the pending set.
Simultaneously, another warp in the pending set will be promoted to the running set.
Here, we assume that a running set includes two warps.
It is very likely that the two warps in a running set come from different thread batches.
Hence, they may compete for different rows in the bank and degrade the row locality.

We can propose a better scheduling policy to improve the row locality, as depicted in Fig.~\ref{fig:tbas}(c):
the running set gathers active warps of the same thread batch as they commonly access the same page (\textit{i.e}., the same row).
If the thread batch in running set does not have sufficient active warps, all the warps of this thread batch are demoted to the pending set, and a new thread batch that has sufficient active warps will be promoted to the running set.

In such a design,  promoting warps may harm the row locality when the rows accessed by the previous active warps and the newly promoted ones are different.
Hence, as shown in Fig.~\ref{fig:tbas}(d), a better promotion scheme can promote a thread batch that is the successor of the demoted thread batch, \textit{e.g.}, promoting (1,0,0) (or (1,1,0)) after demoting (0,0,0) (or (0,1,0)).
Due to the page allocation mechanism, the adjacent thread batch is most likely to access the same row in the bank.

The above sequential thread batch switching often results in a round-robin execution sequence, potentially incurring the burst of memory accesses in a short time. 
As illustrated in Fig.~\ref{fig:tbas}(d), all memory accesses are evoked in the first four scheduling cycles.
The situations that may harm the scheduling efficiency include:
1)~A thread batch demoted by a long operation could access the same page again in the near future. 
However, it may not be scheduled again in time; 
2)~When the thread batches are continuously promoted to the running set, the generated memory-accesses burst is coupled with the lost locality. The prolonged queuing delay in memory controllers may overwhelm the reply network connecting memory controllers and SMs~\cite{bakhoda2010throughput}.

To overcome the above drawbacks, we assign higher promotion priority to older thread batches in the pending set.
We assume the priority of the thread batches in Fig.~\ref{fig:tbas}(a) descends from the left to the right and then from the top to the bottom.
Fig.~\ref{fig:tbas}(e) shows the scheduling sequence of the thread batches considering our proposed promotion priority. 
The improvement of row locality, especially the decreasing of memory access burst, leads to significant reduction in average memory access latency.
We name the scheduling method corresponding to the example presented in Fig.~\ref{fig:tbas}(e) as TBAS.

Besides the maintenance of intra-/inter-thread batch row locality and alleviation on congestion of reply network, TBAS also reduces the stretch of memory access footprint by limiting the active thread batches in a particular time window. Such a limitation on thread-level parallelism can bring in an implicit positive effect on the cache locality~\cite{rogers2012cache} as we shall explain in Section~\ref{subsec:pe}.

The hardware overhead of TBAS is similar to that of CCWS except for the promotion priority arbitrator. 
Fortunately, the number of concurrent thread batches in an SM is usually small: 
An SM of Fermi GPU, for example, supports only eight concurrent thread blocks (or at most 8 thread batches).
Therefore, the implementation overhead of the arbitrator is negligible.

\section{Experiment Methodology}
\label{sec:setup}

\begin{table}[t]
\center
\small
\caption{The characteristics of GPU applications: the category an application belongs to; the used thread organizations labeled as (grid dimension, block dimension).}
\Description[Table 1.]{The characteristics of GPU applications: the category an application belongs to; the used thread organizations labeled as (grid dimension, block dimension).}
{
\vspace{-5pt}
\begin{tabular}{|*{5}{c|}}
\hline \multirow{2}{*}{Application} &  \multirow{2}{*}{Abbreviate} &  \multirow{2}{*}{Category} &  Thread organization &  \multirow{2}{*}{MPKI}\\
&&& (grid dimension, block dimension)&\\
\hline
\hline InvertedIndex &II&C1&(1D, 1D)& 45.23\\
\hline Kmeans Clustering&KM& C1 &(1D/2D, 1D)&11.91\\
\hline PageViewCount&PVC& C1&(1D, 1D)& 5.52\\
\hline PageViewRank &PVR&C1 &(1D, 1D)&2.10\\
\hline Fast Walsh Transform &FWT&C1&(1D, 1D)& 1.01\\
\hline Seven Point Stencil&STEN& C1&(2D, 2D)& 0.85\\
\hline Back Propagation&BP &C1&(1D, 2D)& 0.75\\
\hline Merge Sort&MS& C1&(1D, 1D)& 0.53\\
\hline Discrete Cosine Transform&DCT &C1&(2D, 2D)& 0.37\\
\hline Scalar Products&SP& C1&(1D, 1D)& 0.17\\
\hline Reduction&RD& C1&(1D, 1D)& 0.04\\
\hline Coulombic Potential&CUTCP& C1&(2D, 2D)& 0.03\\
\hline
\hline Neural Network &NN &C2&(1D/2D, 1D)& 1.21\\
\hline Hotspot &HOT&C2&(2D, 2D)& 0.75\\
\hline Sorting Networks&SN &C2 &(1D, 1D)&0.27\\
\hline
\hline Angular Correlation & TPACF& C3&(1D, 1D)& 0.89\\
\hline MUMmerGPU &MUM &C3&(1D, 1D)& 5.85\\
\hline Lattice-Boltzmann Method&LBM& C3&(2D, 1D)& 5.82\\
\hline CFD Solver&CFD& C3&(1D, 1D)& 0.85\\
\hline LIBOR Monte Carlo & LIB &C3&(1D, 1D)& 0.62\\
\hline
\end{tabular}}\label{tbl:app}
\end{table}

\begin{table}
\begin{center}
\small
\caption{The characteristics of CPU applications.}
\Description[Table 2.]{The characteristics of CPU applications.}
{
\begin{tabular}{|c|c|c|c|}
\hline Application&MPKI&Application&MPKI\\
\hline mcf&52.28&bzip2&4.14\\
\hline omnetpp&34.67&h264ref&1.95\\
\hline xalancbmk&27.52&gcc&0.58\\
\hline lbm&20.23&perlbench&0.21\\
\hline
\end{tabular}}\label{tbl:cpuapp}
\end{center}
\end{table}

\begin{table}
	\vspace{-10pt}
\small
\caption{The characteristics of heterogeneous workloads.}
\Description[Table 3.]{The characteristics of heterogeneous workloads.}
\label{tbl:hw}
\begin{minipage}{\columnwidth}
	\begin{center}
	\begin{tabular}{|c|c|c|}
		\hline 
		Workloads& Type \footnote{Each workload type is denoted by the types of combined applications, \emph{e.g.}, NN-I means two memory non-intensive CPU applications run with a memory intensive GPU application.} &Applications \\ \hline
		WL0&NN-N&perlbench, bzip2, FWT\\ \hline 
		WL1&NN-N&gcc, h264ref, BP\\ \hline 
		WL2&NN-I&gcc, bzip2, II\\ \hline 
		WL3&NN-I&perlbench, h264ref, KM\\ \hline 
		WL4&IN-N&omnetpp, gcc, STEN\\ \hline 
		WL5&IN-I&xalancbmk, h264ref, PVC\\ \hline 
		WL6&II-N&mcf, lbm, DCT\\ \hline 
		WL7&II-N&omnetpp, xalancbmk, FWT\\ \hline
		WL8&II-I&mcf, xalancbmk, PVR\\ \hline 
		WL9&II-I&omnetpp, lbm, II\\ \hline 
		WL10&IN-N&lbm, bzip2, NN\\ \hline
		WL11&IN-I&mcf, perlbench, MUM\\ \hline
	\end{tabular}
\end{center}
\end{minipage}
\end{table}
\subsection{Benchmark}
We adopt a set of diverse GPU applications from~\cite{cuda_sdk,bakhoda2009analyzing,che2009rodinia,jog2013owl,stratton2012parboil} as our benchmark used in our evaluations.
Most of the applications are fully simulated except for the applications from~\cite{jog2013owl} of which only the first two billion instructions are simulated.
The detailed characteristics of each application in the benchmark are summarized in Table~\ref{tbl:app}.
All GPU applications are profiled to generate the optimal thread batches before execution.

We combined eight CPU applications with GPU application to construct the heterogeneous workloads for the evaluation.
The CPU workloads are from SPEC CPU 2006, as shown in Table~\ref{tbl:cpuapp}. PinPoint~\cite{luk2005pin} is used to extract the execution phases for all CPU applications. The CPU applications are divided into two types: \textit{memory intensive} where the L2 cache \emph{misses per kilo instructions} (MPKI) is higher than 20; and \textit{memory non-intensive} where the L2 cache MPKI is lower than 20. The GPU applications can be also classified into two types based on L2 cache MPKI -- memory intensive (MPKI$>$2) and non-intensive (MPKI$<$2). Although the L2 cache MPKI of most GPU applications are lower than that of CPU applications, within an arbitrary time window, GPU applications possibly generate two orders of magnitude greater L2 cache misses than CPU applications due to their high instruction throughput (\textit{i.e.}, IPC). Moreover, we grouped GPU application into three categories, C1--C3, according to their sensitivity to TEMP+TBAS (shall be explained in Section~\ref{subsec:pe}). 

We permute the combination of different types of CPU and GPU applications to create twelve heterogeneous workloads. Each workload consists of two CPU applications and one GPU application, as summarized in Table~\ref{tbl:hw}.
We construct ten workloads (WL0--WL9 in Table~\ref{tbl:hw}) where the GPU applications are picked up from C1. Half of GPU applications in WL0--WL9 are memory intensive, while the rest are memory non-intensive. For the CPU workloads in WL0--WL9, we can have three combination types (\textit{i.e.}, NN, IN, and II) of the dual-applications. The generated ten heterogeneous workloads cover most cases where EMU may act variably. We also construct two extra workloads, \textit{i.e.}, WL10 and WL11, each of which consists of one GPU application from C2 and C3, respectively.

\begin{table}
\center
\vspace{-6pt}
\small
\caption{Simulation configuration.}
\Description[Table 4.]{Simulation configuration.}
{
\begin{tabular}{|l|l|}
\hline \multicolumn{2}{|c|}{CPU}\\
\hline Number of Cores&2\\
\hline Execution&3 GHz, OOO 4 issues, 256-entry ROB\\
\hline L1 Data Cache&32KB 4-way, 2-cycle hit, write back, 64B\\
\hline L2 Cache& 2MB 8-way, write back, 64B\\
\hline 
\hline \multicolumn{2}{|c|}{GPU}\\
\hline Number of SMs&8\\
\hline SM Clock&600MHz\\
\hline SIMD width&16\\
\hline L1 Data Cache&32KB 4-way, write through, 128B\\
\hline Warp Size&32\\
\hline Max Number of Threads&1536/SM\\
\hline Max Thread Blocks&8/SM\\
\hline Scheduler&CCWS~\cite{rogers2012cache}, OWL~\cite{jog2013owl}, TBAS\\
\hline TLB&64-entry L1, 8KB page walk cache, 512-entry shared L2\\

\hline \multirow{2}{*}{L2 Cache}& 128B line, 8-way associated, 2 banks,\\
 & 512KB/bank, total 1MB \\
\hline
\hline \multicolumn{2}{|c|}{Shared resources}\\
\hline \# of Memory Channels& 2 for GDDR5, 1 for DDR3\\
\hline Memory Controller (MC)& FR-FCFS~\cite{owens2000memory}, open-page,  64-entry request queue/MC\\
\hline Interconnection& 2D Mesh\\
\hline GDDR5 Banks & 16\\
\hline Timing & from~\cite{gddr5}\\
\hline DDR3 Banks & 8\\
\hline Timing & from~\cite{ddr3}\\
\hline
\end{tabular}}\label{tbl:sm_config}
\end{table}

\subsection{Simulation Platform}
Since the CPU-GPU CC-NUMA has not been shipped by any industrial vendors, we simulate a GPU system attached with a heterogeneous GDDR5-DDR3 DRAM subsystem.
Our system simulation is performed on \textit{gem5-gpu}~\cite{power2015gem5}, and its configuration is listed in Table~\ref{tbl:sm_config}.

The GPU subsystem includes 8 SMs. Each SM has the similar computational capability as the SMs in Fermi and is set to the $\mathrm{600MHz}$ frequency.
The memory bandwidth per \emph{shared-core-clock} is comparable and even higher than that of real high-end heterogeneous processors integrating similar GPU unit~\cite{a10}. 
 As such, we ensure that our platform resembles real product and conducts fair evaluations.

The page size is set to 4KB, a typical size adopted widely.
To avoid the bottleneck of GPGPU TLB and expose the limitation of DRAM bandwidth in heterogeneous shared memory systems, we also optimize the GPU TLB design in our heterogeneous system including per-SM TLB, highly-threaded PTW and shared L2 TLB~\cite{power2014supporting}.
We choose the configuration with CCWS in~\cite{rogers2012cache} as our baseline.

We estimate the GDDR5 DRAM energy consumption through a modified Micron DRAM power calculator~\cite{cal} based on the datasheet~\cite{gddr5}; the DDR3 DRAM energy consumption is directly obtained from Micron DRAM power calculator by feeding the run-time statistics generated from gem5-gpu.

To evaluate the effectiveness of TEMP and TBAS, we compared the following approaches:
\begin{itemize}
	\item \textbf{CCSW} refers to the design for improving the L1 cache locality in GPU proposed by~\cite{rogers2012cache}. The results of CCSW are used as the normalization basis in our evaluations.
		\item \textbf{OWL} denotes the optimized scheduling method proposed by~\cite{jog2013owl}, which improves the performance through optimizing the cache and memory accesses in GPU systems.
	\item \textbf{TEMP} denotes the thread batch enabled memory partitioning scheme presented in Section~\ref{sec:tbmp}.
	\item \textbf{TEMP+TBAS} refers to the design integrating \textit{TEMP} and \textit{TBAS}. 	
	\item \textbf{BW-AWARE} denotes a synergistic bandwidth-aware page placement policy in~\cite{agarwal2015page}. It places the GPU pages across the heterogeneous memory system, {i.e.}, GDDR5 and DDR3 DRAM, and their memory bandwidth is shared across GPU pages.
	\item \textbf{Batching+BW} refers to the scheme that combines TEMP, TBAS, and BW-AWARE. 
\end{itemize}
\section{Result}
\label{sec:re}
\subsection{Evaluation Results for GPU Applications}
\label{subsec:pe}

\begin{figure}[t]
\centering
\includegraphics[width=1\textwidth]{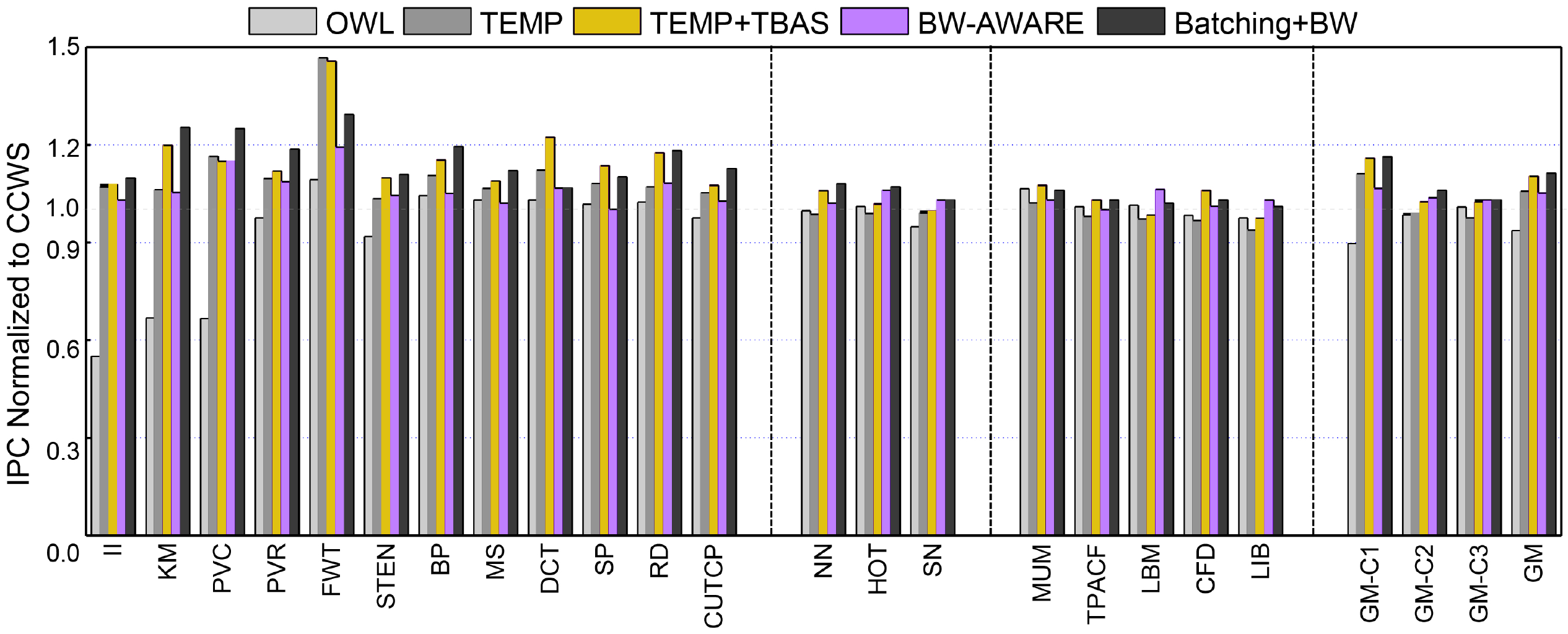}
\caption{The performance (IPC) of different schemes. The results are normalized to CCWS.}
\Description[Fig. 10]{The performance (IPC) of different schemes. The results are normalized to CCWS.}
\label{fig:perf}
\end{figure}

\begin{figure}[t]
	\centering
	\includegraphics[width=1\textwidth]{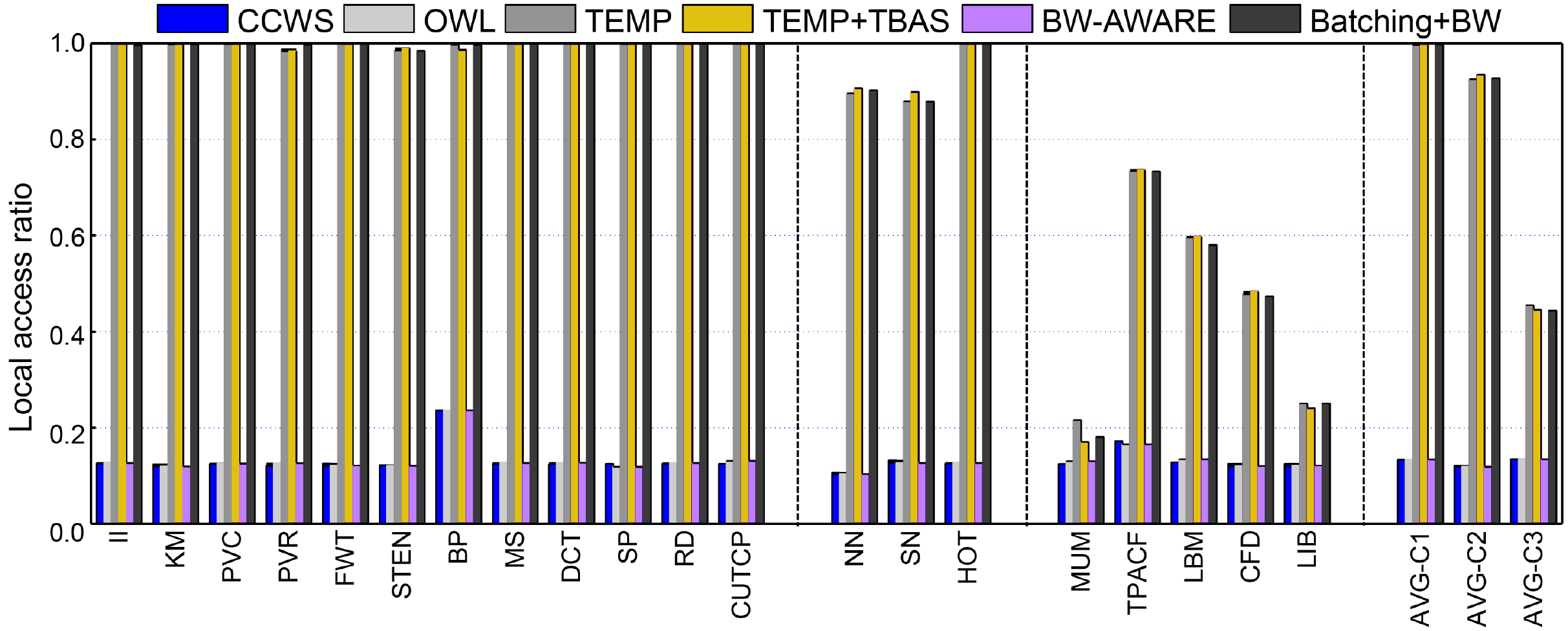}
	\caption{The local access ratio distribution of memory accesses of different schemes.}
	\Description[Fig. 11.]{The local access ratio distribution of memory accesses of different schemes.}
	\label{fig:md}
\end{figure}


\subsubsection{Performance}
We first evaluate and analyze the performance and local access ratio to each memory bank across the different designs for the GPU applications. 
Here, \textit{local access} denotes the memory access from the SM associated with the banks, while \textit{remote access} refers to the access from other SMs.
According to the performance results under the TEMP design and the evaluated local access ratio, GPU applications are classified into the following three categories:
\begin{itemize}
	\item \textbf{C1}:~~These applications present the high local access ratio (on average $>$99\%) and significant performance improvement across all the configurations employed by TEMP. 
	\item \textbf{C2}:~~Similar to C1, the applications in C2 also demonstrate high local access ratio ($>$93\%). 
	In contrast, they present a slight performance reduction ($\sim$1\%) under TEMP yet the effective performance improvement under TEMP+TBAS.
	\item \textbf{C3}:~~
	The applications in C3 do not have high local access ratio due to the intrinsic thread-data mapping and memory access pattern. Their overall performance applied with TBAS and TEMP is degraded compared with those of CCWS.
\end{itemize}
The performance results are shown in Fig.~\ref{fig:perf}, and Fig.~\ref{fig:md} shows the local access ratio for the GPU applications.

\begin{figure}[t]
\centering
\includegraphics[width=.8\textwidth]{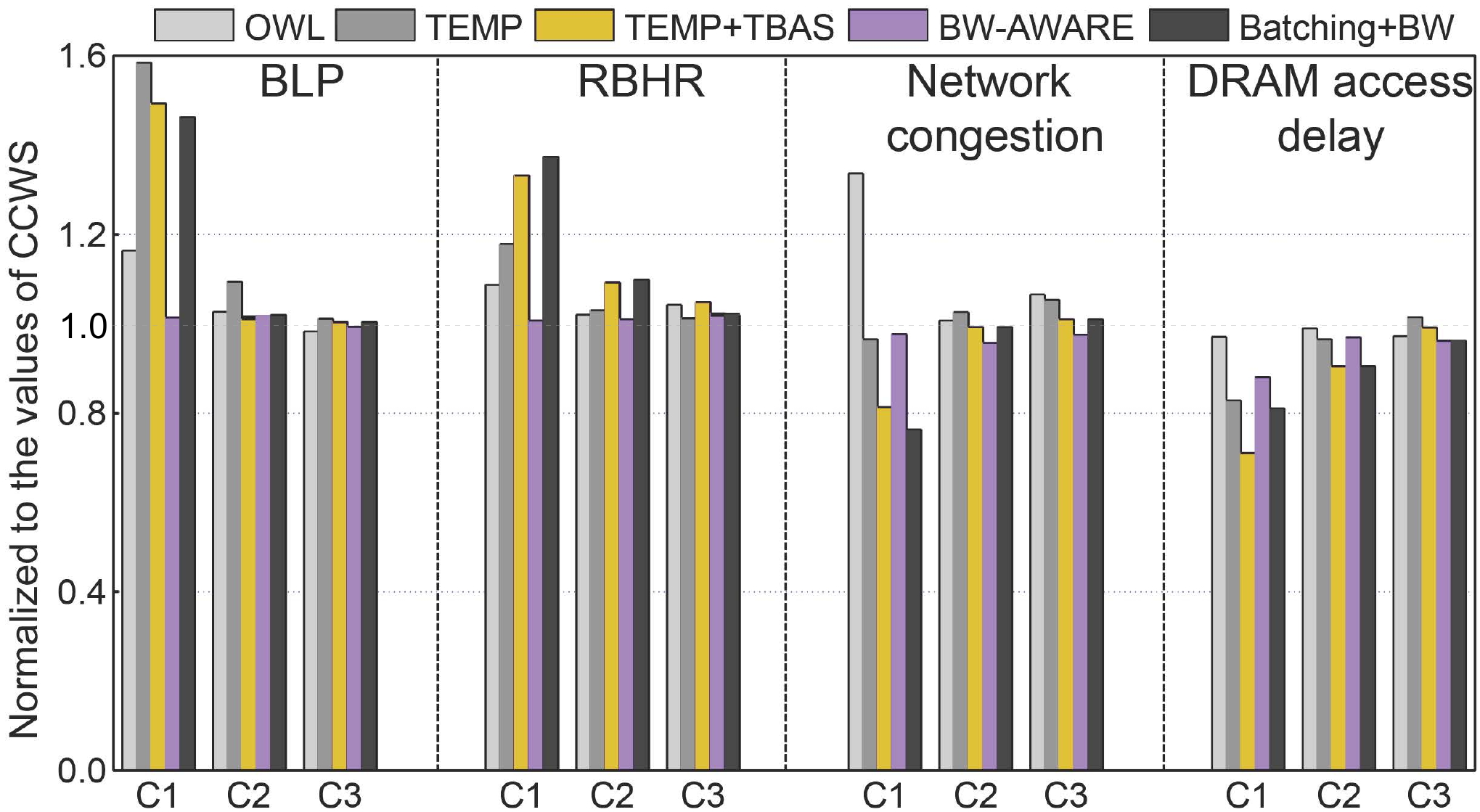}
\caption{The statistic of DRAM system and congestion on reply network.}
\Description{The statistic of DRAM system and congestion on reply network.}
\label{fig:dram_stat}
\end{figure}

The overall results show that applying TEMP on top of CCWS achieves 5.7\% geometric mean (GM) speedup while replacing CCWS with TBAS (\textit{i.e.}, TEMP+TBAS) can further raise the speedup to 10.3\%. 
Based on our evaluations, OWL is 93.6\% within the performance of CCWS across the application workloads.
As shown in Fig.~\ref{fig:md} and Fig.~\ref{fig:dram_stat}, the cache hit rate of OWL is lower than that of CCWS, and the BLP improvement achieved by OWL is limited. 
The results verified that only considering a small subset of thread blocks which share pages is insufficient to achieve remarkable performance improvement.
The IPC of TEMP is 12.9\% higher than that of OWL.
%
BW-AWARE keeps a page placement ratio the same as the bandwidth ratio between GDDR5 and DDR3, which can improve the utilization of the combined bandwidth from both memories. Hence, BW-AWARE gains 5.1\% performance improvement over CCWS as can be seen from Fig.~\ref{fig:perf}. The performance gain is compliance to the value reported in~\cite{agarwal2015page} by given the similar bandwidth ratio.

%
To further evaluate the effects of these designs on the memory requests for the three categories of the GPU applications, we summarize the DRAM usage statistic (BLP, RBHR, DRAM access delay) as well as the \emph{stalls on reply network} connecting memory controllers and SMs induced by the network congestion of these designs.
The results are normalized to those in CCWS and shown in Fig.~\ref{fig:dram_stat}. 

When applying TEMP on C1, the BLP of C1 is significantly improved by 58.3\%, while the RBHR is increased by 17.8\%. As expected, by suppressing the inter-SM interference of memory accesses, TEMP unveils the intrinsic locality and access parallelism of thread batches. In comparison to TEMP, OWL improves BLP by 16.3\% and RBHR by 8.6\%, respectively. 
The opportunistic prefetching adopted by OWL boosts RBHR.
We also investigated the network congestion between the SMs and the GDDR5 DRAM partitions. 
The network congestion of OWL is 33.6\% more than that of CCWS. This value quantitatively demonstrates that CCWS has a higher L1 cache hit rate, less L2 accesses, and less DRAM accesses compared to OWL. 
All the above factors together lead to 17.3\% reduction in DRAM access delay with TEMP in C1. Consequently, TEMP achieves 11.1\% performance improvement over CCWS, which is 24.0\% higher than OWL.
For C1, the BLP in TEMP+TBAS is 9.1\% smaller than that in TEMP. 
This is because the number of active thread batches is intentionally limited for row locality enhancement.
On the other hand, C1's RBHR in TEMP+TBAS is raised by 33.1\% and the DRAM access delay is reduced by 29.9\%. 
More importantly, a considerable reduction in network congestion (18.7\%) is observed. 
As a result, more than 15\% performance improvement is achieved by TEMP+TBAS for C1 as shown in Fig.~\ref{fig:perf}.

C2 achieves a high local access rate when TEMP is applied.
However, TEMP is hard to increase the BLP of C2 since the BLP of C2 already approaches the theoretical upper bound. 
For instance, some kernels in \texttt{NN} have only a few thread blocks whose number is even lower than the bank count.
Applying TEMP on those kernels may limit the BLP. 
Fortunately, TBAS enhances the row locality and reduces the network congestion, resulting in slight speedup ($\sim$2\%). 
As shown in Fig.~\ref{fig:perf}, the performance of C3 in TEMP/TEMP+TBAS is averagely degraded/improved by 2.5\%/2.3\%. 
Note that it is difficult to formalize the thread-data mapping of the applications in C3. 
Thus, applying TEMP for C3 prolongs DRAM access delay.


\begin{figure}[t]
	\centering
	\includegraphics[width=.8\textwidth]{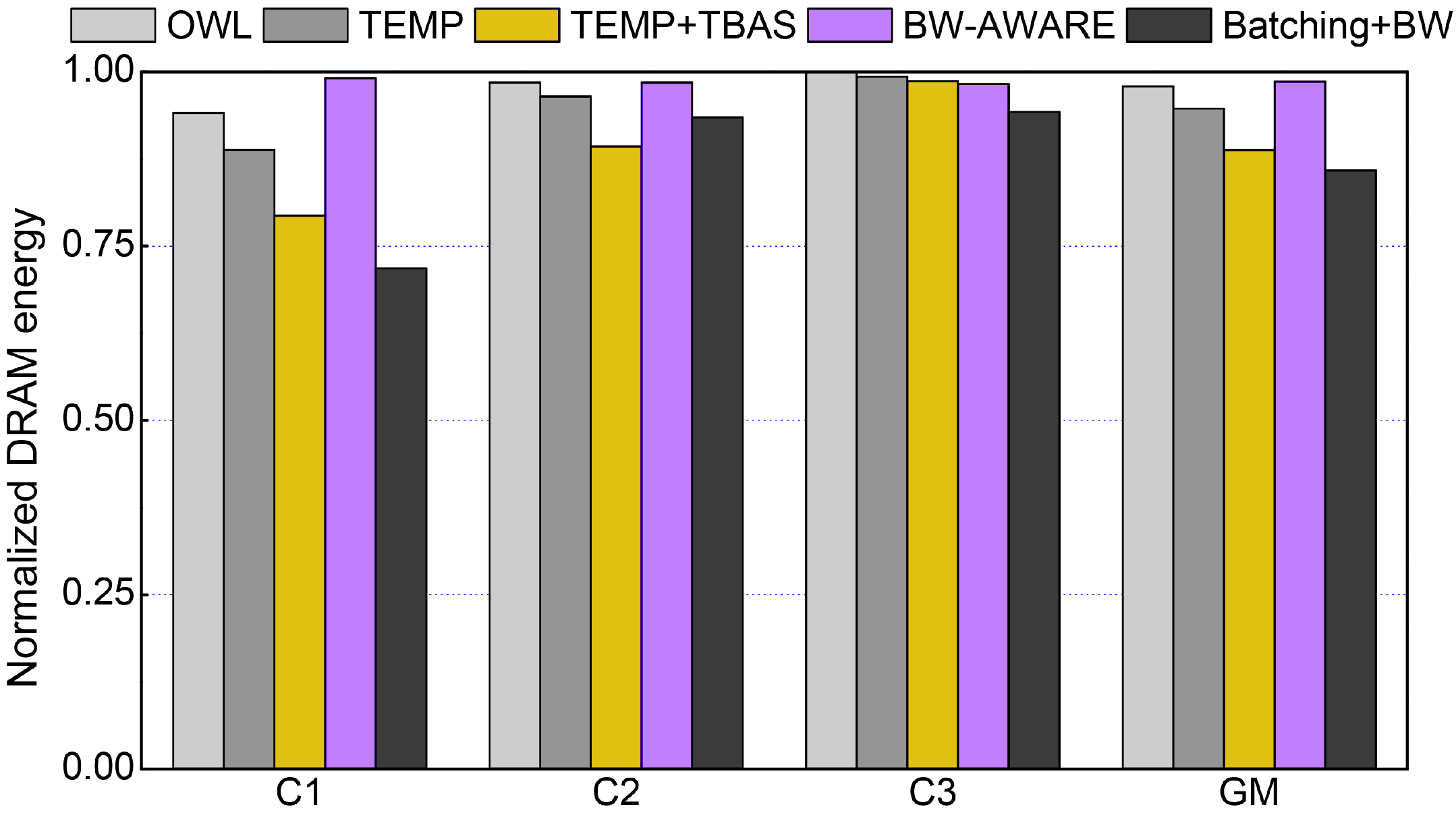}
	\caption{The normalized DRAM energy consumption of different schemes.}
	\Description[Fig. 12.]{The normalized DRAM energy consumption of different schemes.}
	\label{fig:energy}
\end{figure}

\subsubsection{Energy}
The normalized DRAM energy consumption of all configurations is shown in Fig.~\ref{fig:energy}.
Generally, the DRAM energy savings come from two main sources:
1)~The saving of activate energy that dominates DRAM energy consumption, which can be achieved by increasing RBHR; 
and 2)~The saving of the background energy, which is proportional to the reduction of the execution time. 
Therefore, DRAM energy reduction is relevant to the improved access locality as well as the overall performance improvement. 
Our results show that compared to CCWS, the DRAM energy saving of TEMP is 11.2\%. TEMP+TBAS saves 20.7\% more energy than CCWS because of the significantly improved RBHR. OWL saves 5.9\% energy which is less than TEMP+TBAS as the result of the higher row activation ratio and worse performance. Batching+BW achieves the highest energy saving of 14.2\%.

\begin{figure}[t]
\centering
\includegraphics[width=.8\textwidth]{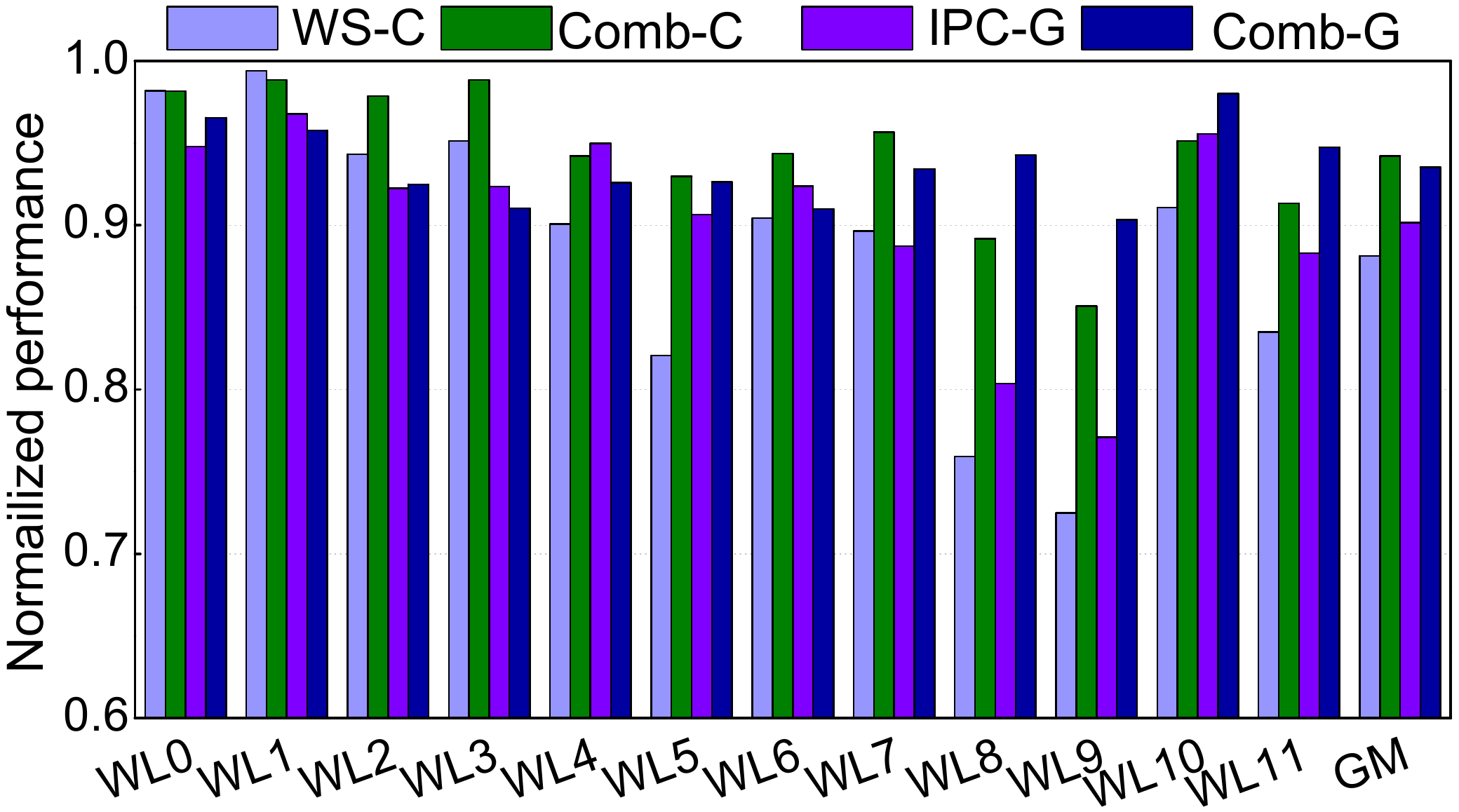}
\caption{The performance of heterogeneous workloads.}
\Description[Fig. 11.]{The performance of heterogeneous workloads.}
\label{fig:heter_ipc}
\end{figure}
\subsection{Evaluation for Heterogeneous Workloads}
Fig.~\ref{fig:heter_ipc} shows the performance of the CPU applications (WS-C) and the GPU application (IPC-G) in each heterogeneous workload when TEMP+TBAS is applied. The performance of the CPU applications in a workload is measured by the weighted speedup~\cite{eyerman2008system}. These results are normalized to the weighted speedup of the same CPU applications running standalone on the heterogeneous system. The IPC of a GPU application is also normalized to the IPC obtained by exclusively running with TEMP+TBAS . 
The memory-intensive CPU and GPU applications in the workloads suffer from non-trivial performance degradation due to the contention for shared resources, e.g., interconnection network and DRAM. On the contrary, the performance degradation of memory non-intensive applications is much less. The weighted performance of CPU applications across twelve workloads is reduced by 11.9\%; correspondingly, the IPC of GPU applications is 9.2\% lower than that obtained by TEMP+TBAS running alone.

The effectiveness of TEMP and TBAS is constrained in the CPU applications and hence, the performance of the CPU applications is degraded as the CPU applications:
1)~TBAS expects consecutive thread blocks to access their physical pages in a limited span of rows. The physical addresses of the pages accessed by the CPU applications, however, can mix with those of the pages accessed by the GPU applications, 
deteriorating the row locality of the GPU applications; 
2)~On the other hand, even if TBAS successfully preserves the row locality of the GPU applications, the memory controller probably always prioritizes the intensive memory accesses from the GPU and suspends the memory accesses from the CPU.

To address the above problems, we can first divide each bank into two portions -- one for CPU and one for GPU. We reserve the rows with higher addresses in a bank for CPU and the ones with lower addresses for GPU.
The new pages for CPU and GPU are from the reserved address space. As such, most pages of CPU and GPU can be physically separated in a bank, which allows TBAS to keep the row locality of GPU applications when CPU applications are running simultaneously. 
Secondly, the memory controller is set to always promote the memory accesses from CPU against the ones from GPU, as proposed in~\cite{ausavarungnirun2012staged}. Since most CPU applications are delay-sensitive, unconditionally promoting the memory accesses from CPU can eliminate the risk of memory access starvation on the CPU-side. 
Combining the above two solutions, the performance loss in the CPU/GPU applications are reduced by 6.1\%/3.5\%, as denoted by Comb-C and Comb-G in Fig.~\ref{fig:heter_ipc}.  We can see that some workloads (e.g., \texttt{WL8} and \texttt{WL9}) including both CPU and GPU intensive applications attain significant performance improvement from the integrated heterogeneous-aware thread batching.

The solutions mentioned above is simple yet capable of keeping the effectiveness of TEMP and TBAS for GPU applications while preventing considerable performance loss for CPU applications. We believe more sophisticated techniques can further balance the throughput between CPU and GPU~\cite{kayiranmanaging}. However, the balanced throughput design is beyond the interests of this paper and left for the future work.

\section{Related Works}
\label{sec:rw}
\subsection{Memory Partitioning in Multi-core Systems}

In multi-core systems, \emph{memory bank partitioning} (MBP) binds a thread to one or more memory banks.
Every thread accesses its own private banks to avoid the interference from other threads.
Mi \emph{et al.}~\cite{mi2010software} first proposed MBP and used modified bank permutation to compensate the degraded BLP.
Jeong \emph{et al.}~\cite{jeong2012balancing} used sub-ranking to overcome the BLP degradation on single thread after applying MBP.
Liu \emph{et al.}~\cite{liu2012software} designed a purely software MBP based on OS page allocation. 
They also explored the utilization of MBP in a multi-threaded application but the result was not very promising because of the inter-thread data sharing.
Xie \emph{et al.}~\cite{xieimproving} pointed out that unbalanced memory requirements across the threads is the main reason of the BLP degradation and then proposed a dynamic bank partitioning approach to solve this problem.
In TBMP, BLP is guaranteed by workload balancing across the SMs while the memory access fairness is guaranteed by the homogeneity of the GPU threads in a kernel.
Thread batching in TEMP also alleviates the negative impact of inter-thread data sharing on system performance in multi-threaded applications.
\subsection{DRAM Efficiency in GPU}
%
Compiler-assisted data layout transformation~\cite{Yang2010,sung2010data,xie2015enabling} proactively prevents unbalanced accesses to DRAM components by carefully allocating the data, register file or the thread block index.
For example, Xie \emph{et al.} {\cite{xie2015enabling}} put forward a compiler-based framework to balance the register allocation and the targeted thread-level parallelism in the GPU system.
However, the compiler-level methods are not aware of any hardware implementation details. Both thread scheduling and DRAM address mapping at the hardware level may offset the optimization brought by the compiler level. The hardware-level approaches of enhancing DRAM usage efficiency in GPU or CPU-GPU systems include:

\textit{Enhanced memory schedulers:}~~Jeong \emph{et al.}~\cite{jeong2012qos} designed a QoS-aware memory scheduler for MPSoC with CPUs and GPUs. The DRAM bandwidth allocation between the CPUs and GPUs is dynamically adjusted to meet the frame rate requirement of the GPUs and maximize the overall system throughput. Ausavarungnirun \emph{et al.}~\cite{ausavarungnirun2012staged} proposed a staged memory scheduling framework with affordable hardware cost for heterogeneous systems. We adopt the memory scheduling policy from~\cite{ausavarungnirun2012staged} to customize our proposed heterogeneous-aware thread batching.
    
\textit{Enhanced thread scheduler:}~~Jog \emph{et al.}~\cite{jog2013owl} revealed that serial thread block data layout and sequential thread block dispatching can cause BLP degradation of GPU applications. A scheduler is then designed to improve the BLP by prioritizing the thread blocks in consecutive SMs.
    The authors also utilized prefetching to compensate the degradation of row locality.
    However, if the memory of a GPU is pageable, the effect of prioritized thread scheduling will become uncertain, because the pages of consecutive thread blocks can be nonconsecutive or not concentrated to a DRAM row.
    In our scheme, TEMP relies on thread batching and page coloring to improve the BLP and TBAS enhances the row locality, targeting a heterogeneous system design supporting pageable GPU memory.
\section{Conclusion}
\label{sec:cc}
Modern GPUs suffer from the mismatching between thread-level parallelism and DRAM bandwidth.
To improve the DRAM usage efficiency of GPU applications, we propose an integrated architectural approach which is composed of TEMP and TBAS techniques:
TEMP improves memory access parallelism for massive multi-threaded GPU applications by minimizing the memory access interweaving
across SMs;
and TBAS maximizes the row locality by elaborately prioritizing the execution of the thread batches. Heterogeneous-aware thread batching is also introduced to promise the effectiveness of thread batching when running heterogeneous workloads.
Our results show that TEMP+TBAS can achieve up to 10.3\% system performance improvement and 11.3\% DRAM energy saving 
compared to the baseline employing CCWS. By using the simple and existing solution, the heterogeneous-aware thread batching can still maintain 93.9\% CPU performance and 96.5\% GPU performance compared to the results of exclusively running CPU and GPU applications.
\begin{acks}
This work is supported in part by US National Science Foundation under Grant 1725456 and Grant 1615475; Bing Li acknowledges the National Academy of Sciences (NAS), USA for awarding the NRC research fellowship. 
\end{acks}
%
\bibliographystyle{ACM-Reference-Format}
\bibliography{9references}

\end{document}